# Tycho Brahe's 1572 supernova as a standard type Ia explosion revealed from its light echo spectrum

Oliver Krause[1], Masaomi Tanaka[2,3], Tomonori Usuda[4], Takashi Hattori[4], Miwa Goto[1], Stephan Birkmann[1,5], Ken'ichi Nomoto[2,3]

[1]*Max-Planck-Institut für Astronomie, Königstuhl 17, 69117 Heidelberg, Germany*

[2]*Institute for the Physics and Mathematics of the Universe, University of Tokyo, Kashiwanoha 5-1-5, Kashiwa, Chiba 277-8568, Japan*

[3]*Department of Astronomy, Graduate School of Science, University of Tokyo, Hongo 7-3-1, Bunkyo-ku, Tokyo 113-0033, Japan*

[4]*SUBARU Telescope, National Astronomical Observatory of Japan, 650 North A'ohoku Place, Hilo, Hawaii, USA*

[5]*European Space Agency, Space Science Department, Keplerlaan 1, 2200 AG Noordwijk, The Netherlands*

**Type Ia supernovae (SNe Ia) are thermonuclear explosions of white dwarf stars in close binary systems[1]. They play an important role as cosmological distance indicators and have led to the discovery of the accelerated expansion of the Universe[2,3]. Among the most important unsolved questions[4] are how the explosion actually proceeds and whether accretion occurs from a companion or via the merging of two white dwarfs. Tycho Brahe's supernova of 1572 (SN 1572) is thought to be one of the best candidates for a SN Ia in the Milky Way[5]. The proximity of the SN 1572 remnant has allowed detailed studies, such as the possible identification of the binary companion[6], and provides a unique**

**opportunity to test theories of the explosion mechanism and the nature of the progenitor. The determination of the yet unknown[7,8,9] exact spectroscopic type of SN 1572 is crucial to relate these results to the diverse population of SNe Ia[10]. Here we report an optical spectrum of Tycho Brahe's supernova near maximum brightness, obtained from a scattered-light echo more than four centuries after the direct light of the explosion swept past Earth. We find that SN 1572 belongs to the majority class of normal SNe Ia. The presence of a strong Ca II IR feature at velocities exceeding 20,000 km s$^{-1}$, which is similar to the previously observed polarized features in other SNe Ia, suggests asphericity in SN 1572.**

The supernova of 1572 marked a milestone in the history of science. Danish astronomer Tycho Brahe concluded from his accurate observations of the "new star" in the constellation of Cassiopeia that it must be located far beyond the Moon[11]. This contradiction to the Aristotelian concept, that a change on the sky can only occur in the sub-lunar regime, ultimately led to abandoning the immutability of the heavens. The classification of SN 1572 has been controversial. On the basis of historical records of the light curve and colour evolution, SN 1572 has been interpreted as SN Ia of either a normal or somewhat overluminous type[7] or of a subluminous type[8]. Core-collapse supernovae of types Ib[9] or II-L[12] have also been suggested to be compatible with the light curve. Support for a SN Ia has been inferred from X-ray studies of the ejecta composition[13], but the determination of the exact supernova type has not been possible without spectroscopic information.

The discovery of light echoes from historic Galactic supernovae, due to both scattering and absorption/re-emission of the outgoing supernova flash by the interstellar dust near the remnant[14,15], raised the opportunity to conduct spectroscopic post-mortems of historic Galactic supernovae. Such a precise determination of the spectral type long after the original explosion has recently been performed for the Cassiopeia A supernova[16] and a supernova in the Large Magellanic Cloud[17].

We obtained Johnson R-band images of the recently identified[15] light echo fields of SN 1572 using the 2.2m and 3.5m telescopes at Calar Alto, Spain, on 23 August 2008 and 2 September 2008. One of the fields observed on 2 September 2008, at an angular distance of d = 3.15° and position angle 62° from the SN 1572 remnant, showed a bright and extended nebulosity with a peak surface brightness of R = 23.6 +/- 0.2 mag arcsec$^{-2}$ (Fig. 1a). The position of this new emission feature, relative to the previously reported light echo detection, has shifted away from the SN 1572 remnant.

The region was re-observed using the FOCAS instrument at the Subaru 8.2m telescope on Mauna Kea, Hawaii, on 24 September 2008 (Fig. 1b). The peak of the emission with a surface brightness of R = 23.5 +/- 0.2 mag arcsec$^{-2}$ has shifted again away from SN 1572. The 1.4 +/- 0.2 arcsec shift within 22 days is consistent with a light echo origin. A long-slit spectrum of the brightness peak of the echo structure at position RA 00h52m12.79s; Dec +65°28'49".7 (J2000.0) was obtained with FOCAS in the same night, covering the wavelength range from 3,800 to 9,200 Å with a spectral resolution of 24 Å.

The acquired echo spectrum unambiguously shows light of a supernova origin (Fig. 2): A number of broad absorption and emission features from neutral and singly ionized intermediate mass elements are detected, all of which are commonly observed in supernovae[10,18]. Type I supernovae are distinguished from those of type II by the absence of hydrogen, and SNe Ia are further distinguished from type Ib and type Ic supernovae by a prominent Si II 6,355 Å absorption feature at maximum light. This feature is clearly seen in the SN 1572 spectrum as a deep absorption minimum at 6,130 Å with a width of 9,000 km s$^{-1}$ at half maximum. The absorption minimum of the line corresponds to a velocity of 12,000 km s$^{-1}$, a typical velocity for normal SNe Ia at maximum brightness[10,18]. Other strong features detected in the spectrum are Si II 4,135 Å, Fe II, Fe III, Na I D + Si II 5,972 Å, O I 7,774 Å and the Ca II infrared triplet.

The echo spectrum represents supernova light over an interval of time around maximum brightness being averaged through the spatial extent of the scattering cloud. We therefore compared the echo spectrum with the spectra of other SN Ia time averaged over the brightness peak of the light curve from 0 to 90 days after explosion. The light echo spectrum of SN 1572 matches such comparison spectra of four well observed normal SNe Ia (1994D, 1996X, 1998bu, 2005cf) and a SN Ia composite spectrum[19] remarkably well. Even faint notches observed in normal SN Ia spectra at 4,550, 4,650 and 5,150 Å[18] can be recognized. Values of the reduced chi-squared ($\chi^2$) from the comparison range between $\chi^2 = 1.5$ and $\chi^2 = 2.5$. The agreement of the spectra indicates that the scattering dust cloud is homogeneous on a length scale of at least 90 light days.

We have compared the spectrum of SN 1572 with thermonuclear supernovae of different luminosity. Both sub- and overluminous SNe Ia, such as SN 1991bg[20] and SN 1991T[21] respectively, showed peculiarities in their spectra near maximum light. SN 1991T lacked a well-defined Si II 6,355 Å absorption feature at maximum light, though the subsequent evolution was similar to normal SNe Ia. The lack of Si II absorption is visible as an imprint in the time-averaged spectrum and different to the strong Si II feature in our spectrum of SN 1572. The class of subluminous objects shows a characteristic deep absorption trough at a wavelength of 4,200 Å, attributed to Ti II[20], near maximum light. Such a feature is not seen for SN 1572. The sub- and overluminous SNe Ia templates, obtained in the same way as described by ref. 19, do not provide a good match to our spectrum of SN 1572 with values of $\chi^2 = 8.6$ and $\chi^2 = 10.1$, respectively (Fig. 3).

A well established correlation between the measured decline $\Delta m15(B)$ of the supernova B-band brightness at maximum and 15 days later has been successfully used to calibrate the absolute brightness of SNe Ia[22]. Applying the range of $1.0 < \Delta m15(B) < 1.3$ obtained from the four comparison SNe Ia yields an absolute brightness of $M_V = -19.0$ +/- 0.3 mag ($H_0 = 72$ km s$^{-1}$ Mpc$^{-1}$). Historical records indicate a maximum brightness between $m_V = -4.0$ mag and $m_V = -4.5$ mag. Accounting for interstellar foreground extinction ($Av = 1.86$ +/- 0.2 mag[8]) this places SN 1572 at a distance of $3.8^{+1.5}_{-0.9}$ kpc. This is larger than the currently most quoted range between 2.3 and 2.8 kpc. However, it is interesting to note that 3.8 kpc is consistent with the non-detection of the remnant at γ-ray energies[23] and the distance to the reported surviving binary companion of SN 1572[24]. If this discovery of the companion is confirmed it would provide evidence for a single-degenerate scenario.

An interesting difference between the spectra of SN 1572 and those of normal SNe Ia is a deep absorption feature at 7,980 Å in the vicinity of the Ca II infrared triplet (Fig. 2). While normal SNe Ia, including SN 1572, show a photospheric absorption minimum of the Ca II infrared triplet at a velocity of 13,500 km s$^{-1}$, the additional sharp absorption in SN 1572 corresponds to a high velocity component of the Ca II triplet at a velocity of 20,000 – 24,000 km s$^{-1}$. Further absorption is detected up to a velocity of 30,000 km s$^{-1}$. High velocity (HV) components have been shown to be ubiquitous in early spectra of SNe Ia[25], however they are often mixed with the photospheric absorption. To our knowledge, a feature as strong as that observed in the SN 1572 spectrum has only occasionally been detected in SN Ia spectra, e.g. in SN 2001el. Its spectrum showed a HV Ca II feature at a velocity of 20,000 - 26,000 km s$^{-1}$ that was kinematically distinct from the photospheric Ca II absorption[26,27]. For SN 2001el spectropolarimetric observations[26,28] have demonstrated that the HV Ca II feature is the result of an aspherical explosion. Thus the similarity of the Ca II HV features of SNe 1572 and 2001el suggests that the asphericity of SN 1572 is also similar to that of SN 2001el. The asphericity could be either due to accretion from a companion or an effect of the explosion.

An exciting opportunity would be to use other SN 1572 light echo spectra, in different spatial directions, to construct a 3-dimensional spectroscopic view of the explosion. This will enable us to determine from observations of a single supernova to what extent spectroscopic diversity can be readily explained by pure geometry effects. Such observations will further constrain the asphericity suggested by our spectrum. In case the HV Ca II feature is caused by a single clump[28], no such feature is expected to be observed in different space directions. Alternatively, the HV Ca II feature could be caused by the interaction with the circumstellar disk of the binary progenitor, in which case the feature might be observed also in other directions[29].

An intriguing possibility to determine an independent geometrical distance to the SN 1572 remnant is to search for echoes with the highest degree of polarization and measure their angular distance from the remnant, as recently performed for other echoes around a Galactic variable star[30]. Since the linear polarization of the scattered light is highest for a scattering angle of 90°, such echoes must be located at a linear distance of $c \times t$ from the SN 1572 remnant, where $c$ corresponds to the speed of light and $t$ to the time since the original explosion.

# References

[1] Nomoto, K., Thielemann, F.-K. & Yokoi, K. Accreting white dwarf models of Type I supernovae. III - Carbon deflagration supernovae. *Astrophys J.* **286**, 644-658 (1984)

[2] Riess, A.G. et al. Observational Evidence from Supernovae for an Accelerating Universe and a Cosmological Constant. *Astron. J.* **116**,1009-1038 (1998)

[3] Perlmutter, S. et al. Measurements of Omega and Lambda from 42 High-Redshift Supernovae. *Astrophys. J.* **517**, 565-586 (1999)

[4] Hillebrandt, W. & Niemeyer,J.C. Type Ia Supernova Explosion Models. *Annu. Rev. Astron. Astrophys*. **38,** 191-230 (2000)

[5] Baade, W. B Cassiopeiae as a supernova of Type I. *Astrophys. J.* **102**, 309–317 (1945)

[6] Ruiz-Lapuente, P. et al. The binary progenitor of Tycho Brahe's 1572 supernova *Nature* **431**, 1069-1072 (2004)

[7] Ruiz-Lapuente, P. Tycho Brahe's supernova: light from centuries past *Astrophys. J.* **612**, 357–363 (2004).

[8] van den Bergh, S. Was Tycho's supernova a subluminous supernova of type Ia? *Astrophys J.* **413**, 67-69 (1993)

[9] Schaefer, B. The Peak Brightnesses of Historical Supernovae and the Hubble Constant. *Astrophys. J.* **459**, 438-454 (1996)

[10] Branch, D. et al. Comparative Direct Analysis of Type Ia Supernova Spectra. II. Maximum Light. *PASP* **118**, 560-571 (2006)

[11] Brahe, T. Astronomiae Instauratae Progymnasmata. In Opera omnia Vol 2 (ed. Dreyer, I.L.E.) 307 (Swets & Zeitlinger, Amsterdam, 1972) (1603).

[12] Doggett, J. B., Branch, D. A comparative study of supernova light curves. *Astron. J.* **90**, 2303-2311 (1985)

[13] Decourchelle, A. et al. XMM-Newton observations of the Tycho supernova remnant. *Astronomy & Astrophysics* **365**, L218-L224 (2001)

[14] Krause, O. et al. Infrared Echoes near the Supernova Remnant Cas A. *Science* **308**, 1604-1606 (2005)

[15] Rest, A. et al. Scattered-Light Echoes from the Historical Galactic Supernovae Cassiopeia A and Tycho (SN 1572). *Astrophys. J.* **681**, L81-L84 (2008)

[16] Krause, O. et al. The Cassiopeia A Supernova was of Type IIb. *Science* **320**, 1195-1197 (2008)

[17] Rest, A. et al. Spectral Identification of an ancient supernova using light echoes in the Large Magellanic Cloud. *Astrophys. J.* **680**, 1137-1148 (2008)

[18] Filippenko, A. Optical Spectra of Supernovae. *Annu. Rev. Astron. Astrophys.* **35**, 309-355 (1997)

[19] Nugent, P., Kim, A. & Perlmutter, S. K-Corrections and Extinction Corrections for Type Ia Supernovae. *PASP* **114**, 803-819 (2002); template spectra available at http://www.supernova.lbl.gov/nugent/spectra.html

[20] Filippenko, A. et al. The subluminous, spectroscopically peculiar type IA supernova 1991bg in the elliptical galaxy NGC 4374. *Astron. J.* **104**, 1543-1556 (1992)

[21] Filippenko, A. et al. The peculiar Type IA SN 1991T - Detonation of a white dwarf? *Astrophys. J.* **384**, L15-L18 (1992)

[22] Phillips,M. M. et al. The Reddening-Free Decline rate Versus Luminosity Relationship for Type Ia Supernovae. *Astron. J.* **118,** 1766-1776 (1999)

[23] Völk, H. et al. Internal dynamics and particle acceleration in Tycho's SNR. *Astronomy & Astrophysics* **483**, 529-535 (2008)

[24] Gonzalez Hernandez, J. et al. The Chemical Abundances of Tycho G in Supernova Remnant 1572. *Astrophys. J.* in the press. arXiv:0809.0601 (2008)

[25] Mazzali, P.A. et al. High-Velocity Features: A Ubiquitous Property of Type Ia Supernovae. *Astrophys. J.* **623**, L37-L40 (2005)

[26] Wang, L. et al. Spectropolarimetry of SN 2001el in NGC 1448: Asphericity of a Normal Type Ia Supernova. *Astrophys. J.* **591**, 1110-1128 (2003)

[27] Mattila, S. et al. Early and late time VLT spectroscopy of SN 2001el – progenitor constraints for a type Ia supernova. *Astronomy & Astrophysics* **443**, 649-662 (2003)

[28] Kasen et al. Analysis of the Flux and Polarization Spectra of the Type Ia Supernova SN 2001el: Exploring the Geometry of the High-Velocity Ejecta. *Astrophys. J.* **593**, 788-808 (2003)

[29] Tanaka, M. et al. Three-dimensional Models for High-Velocity Features in Type Ia Supernovae. *Astrophys. J.* **645**, 470-479 (2006)

[30] Sparks, W. B. et al. V838 Monocerotis: a Geometric Distance from Hubble Space Telescope Polarimetric Imaging of its Light Echo. *Astron. J.* **135**, 605-617 (2008)

**Acknowledgements** This work is based on data collected at Subaru Telescope, which is operated by the National Astronomical Observatory of Japan, and the German-Spanish Astronomical Center, Calar Alto, jointly operated by the Max-Planck-Institut für Astronomie Heidelberg and the Instituto de Astrofísica de Andalucía (CSIC). We thank Ulrich Thiele and the Calar Alto observers for their support. MT is supported by the JSPS (Japan Society for the Promotion of Science) Research Fellowship for Young Scientists. This research has been supported in part by World Premier International Research Center Initiative (WPI Initiative), MEXT, Japan.

**Competing interests statement** The authors declare that they have no competing financial interests.

**Correspondence** and requests for materials should be addressed to Oliver Krause (krause@mpia.de) or Ken'ichi Nomoto (nomoto@astron.s.u-tokyo.ac.jp)

# Captions

Figure 1. Optical images of the SN 1572 light echo. Panels a) and b) show R-band images of the same area of 120 x 120 arcsec². The corresponding observing epochs are labelled. The position of the brightness peak in the first epoch is marked for reference (red cross). The rectangle shown in a) indicates the location of a previous light echo detection[15]. The vector towards the remnant of SN 1572 is indicated (arrow). The seeing for panels a) and b) was 1.5 and 0.9 arcsec, FWHM, respectively. Integration times of the two images were 20 and 12 min. Image reduction was performed using standard methods with IRAF.

Figure 2. Spectrum of SN 1572 and comparison spectra of normal SNe Ia. Important spectral features are marked. The spectra are plotted logarithmically in flux units and shifted for clarity. The spectrum was obtained with two grisms in the blue (150+L550 for λ<5000Å) and red (300B+Y47 for λ>5000Å). Total integration time was 4 hours; 2.5 h for red and 1.5 h for blue channel. The spectrum was extracted from a 2.8 arcsec x 2.0 arcsec aperture (position angle PA = 81°) positioned at the echo brightness peak and then binned to a resolution of 11.2 Å pixel$^{-1}$ and smoothed by a moving average over 5 pixels. Flux calibration was performed against the standard star G191B2B that was observed at comparable airmass. The uncertainty of the flux calibration is 15%. Atmospheric A-band and B-band absorptions were removed using the stellar spectrum of a K star observed in the same slit as the echo. The spectrum has then been corrected for the colour dependence of the scattering process for a scattering angle of $\theta$ = 84° and de-reddened for a foreground extinction of $A_V$ = 4.2 mag. The scattering angle of $\theta$ = 84° results from the light echo geometry: Since all echo emission at a given epoch is located on an ellipsoidal sphere with the Earth and SN 1572 in its foci the echo geometry can be accurately determined. For a distance range to the Tycho remnant of 2.3 – 2.8 kpc the distance and scattering angle of the echo knot are $d$ = 460 +/- 45 light yr and $\theta$ = 90 +/- 5° respectively. For a larger distance of 3.8 kpc the scattering angle is smaller, $\theta$ = 67°, leading to a slightly redder corrected spectrum. However, a slight increase of the adopted foreground extinction by $\Delta A_v$ = 0.08 mag compensates for this effect. The comparison spectra has been obtained from the time average of light curve weighted spectra at days (relative to maximum brightness) -5, -4, -2, +2, +4, +10, +11, +24, +50, +76 for SN 1994D and days -9, -4, +1, +9, +18, +40 for SN 2001el[26,27] and from days -20 through 70 for the normal SN Ia template[19].

Figure 3. Comparison of SN 1572 with SNe Ia of different luminosities. Spectral templates of subluminous, normal and overluminous type Ia SNe are shown in comparison with the spectrum of SN 1572. The spectrum of SN 1572 has been corrected for a scattering angle of $\theta$ = 84° and a foreground extinction of $A_V$ = 4.2 mag as described for Fig. 2. The comparison spectra have been derived as the time average of a spectral series[19] over days 0-90 after explosion and scaled to the spectrum of SN 1572. Specific features typical for the three subtypes are indicated. For the comparison with the intrinsically redder subluminous template, the spectrum SN 1572 was de-reddened for $A_V$ = 3.9 mag.

**Figure 1**

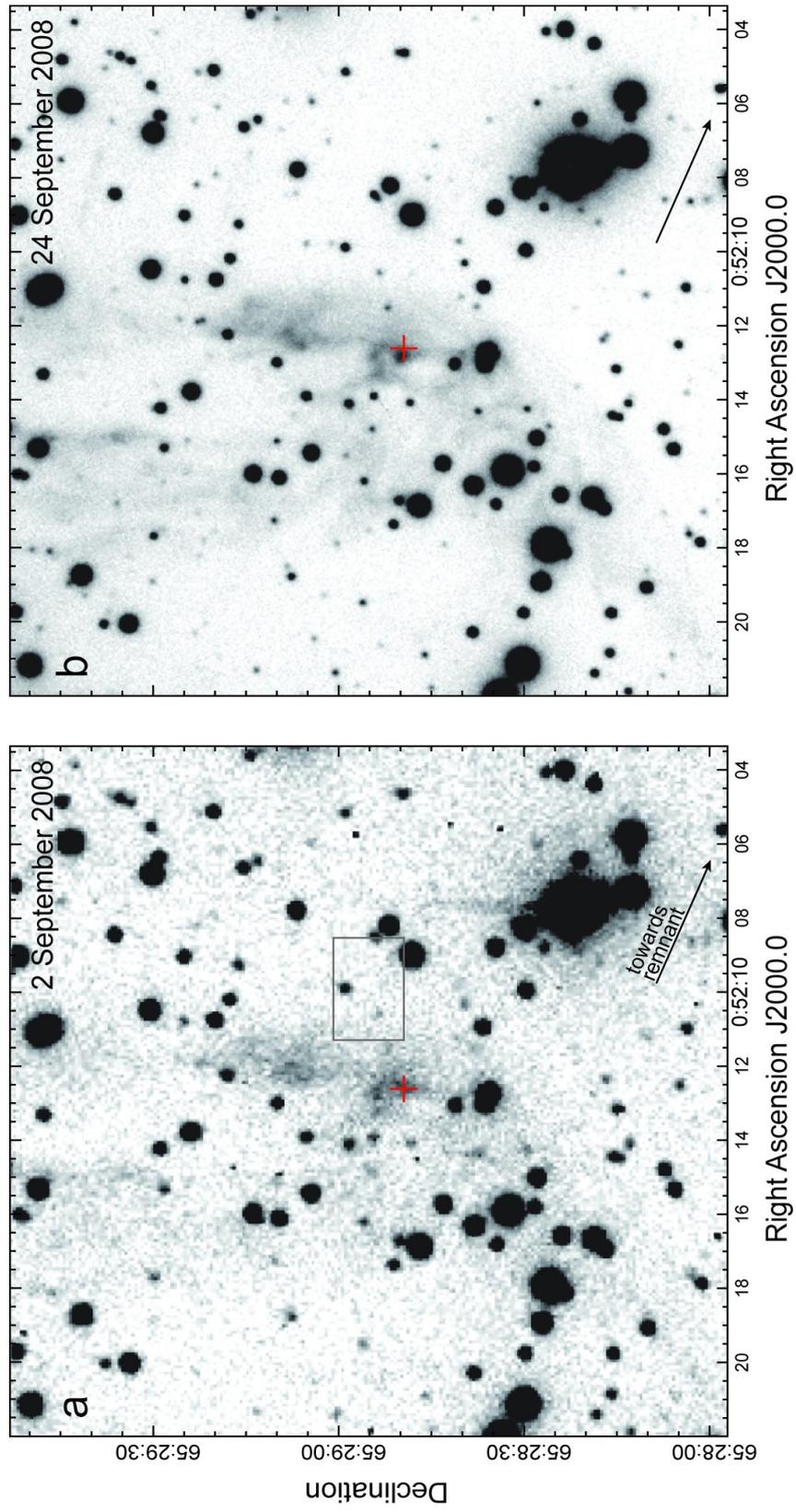

**Figure 2**

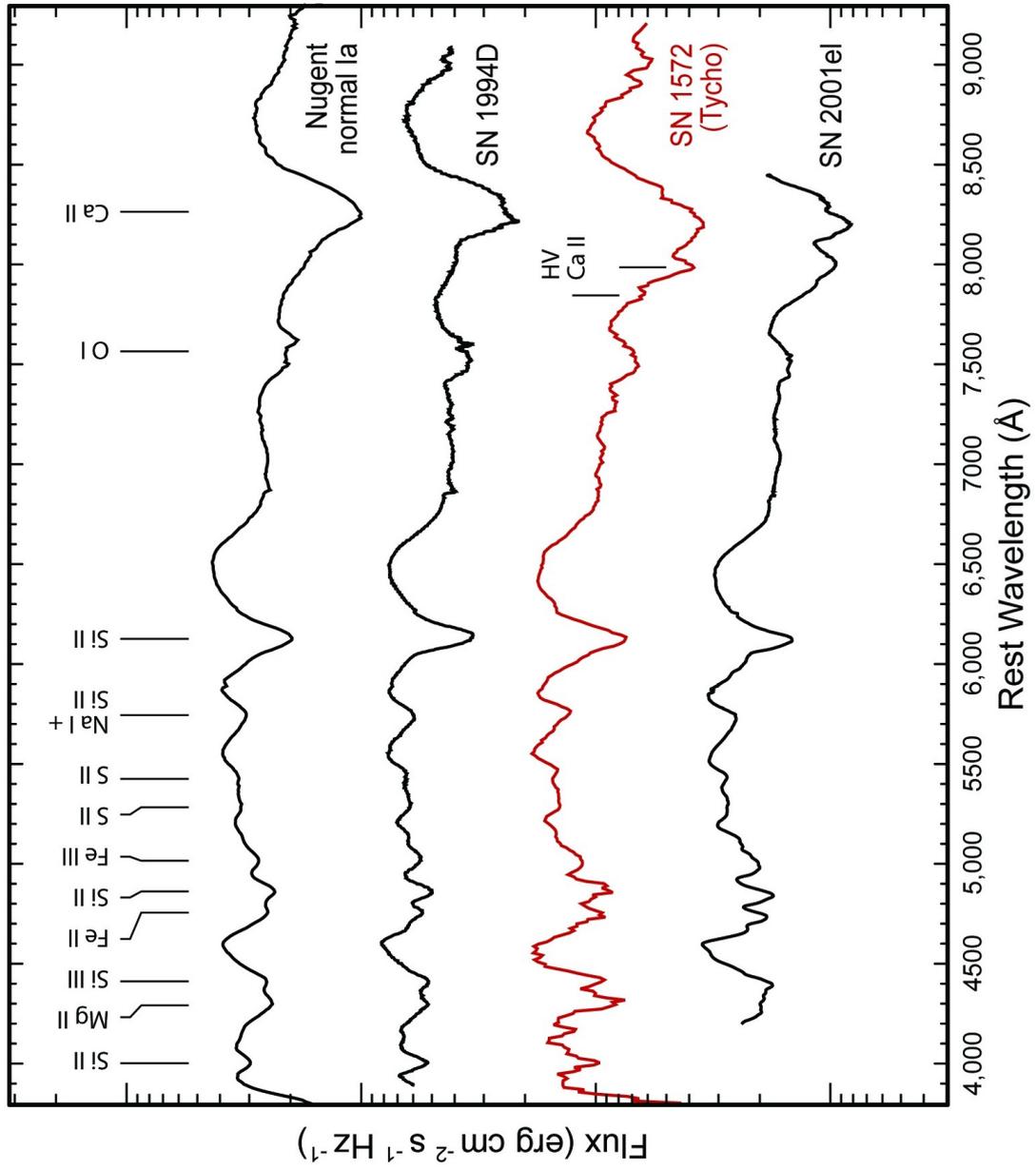

**Figure 3**

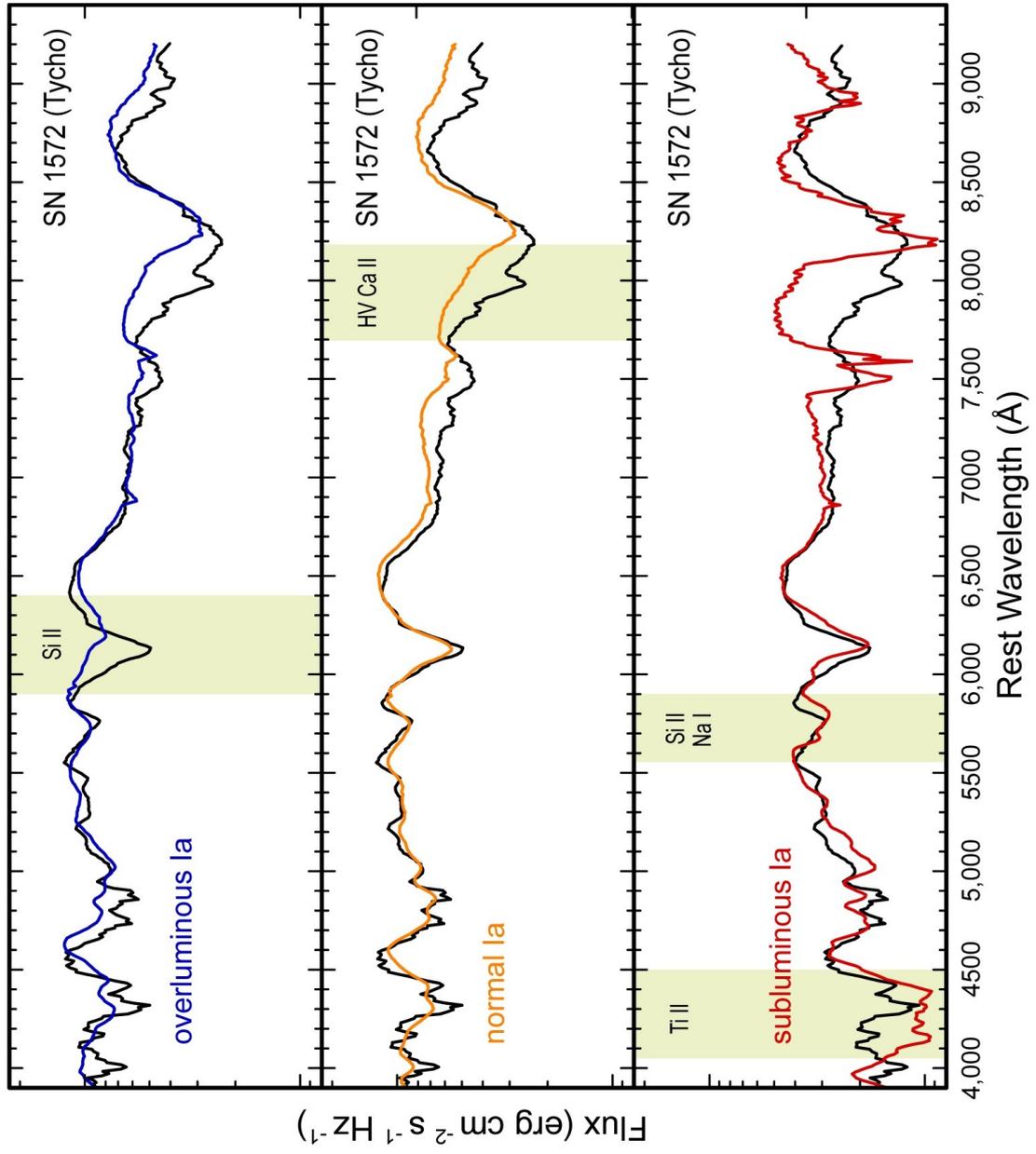